\documentclass[aps,pra,superscriptaddress,amsmath,amsfonts,amssymb,twocolumn]{revtex4-2}

\usepackage{graphicx}
\usepackage{hyperref}
\usepackage{amsthm}
\usepackage{amsfonts}

\newcommand{\uba}{Departamento de F\'\i sica, FCEyN, UBA, Pabell\'on 1,
  Ciudad Universitaria, 1428 Buenos Aires, Argentina}
\newcommand{\ifiba}{Instituto de F\'\i sica de Buenos Aires, UBA CONICET,
  Pabell\'on 1, Ciudad Universitaria, 1428 Buenos Aires, Argentina}

\newcommand{\Tr}{\text{tr}}

\newcommand{\ketbra}[2]{|{#1}\rangle\langle{#2}|}

\newcommand{\comm}[2]{\left[{#1},{#2}\right]}

\newcommand{\beq}{\begin{equation}}
\newcommand{\eeq}{\end{equation}}
\newcommand{\bse}{\begin{subequations}}
\newcommand{\ese}{\end{subequations}}
\newcommand{\bea}{\begin{eqnarray}}
\newcommand{\eea}{\end{eqnarray}}

\usepackage{xcolor}

\begin{document}

\title{Collective effects and quantum coherence in dissipative charging of quantum batteries}

\author{Franco Mayo}
\email{fmayo@df.uba.ar}
 \affiliation{\uba} \affiliation{\ifiba}
\author{Augusto J. Roncaglia}
  \email{augusto@df.uba.ar}
  \affiliation{\uba} \affiliation{\ifiba}


\begin{abstract}
    We consider the dissipative charging process of quantum batteries in terms of a collisional model, where the batteries are coupled to a heat bath using non-energy preserving interactions. First, we show that for low temperatures the collective  process
    can attain a charging power that increases polynomially with the number of batteries. The scaling we find is $N^3$ that, while being grater than the bound obtained for unitary processes, it has a lower efficiency. Then, we study the dissipative charging process of single battery using a time dependent Hamiltonian that generates coherences in the energy basis. In this case we find that the presence of coherence could enhance the charging power and also its efficiency. Finally, we show how this process can be used in a quantum heat engine that contains the charging process as one of its open strokes. 
\end{abstract}

\maketitle

\section{Introduction}

Quantum batteries are devices that allow to temporarily store energy. This concept was coined by Alicki and Fannes~\cite{alicki2013entanglement} when considering whether quantum effects could be used to improve the amount of work that can be extracted from a quantum system using unitary operations. Remarkably, it was soon realized that collective operations allows to extract more work from these devices and  they also allow to speedup their power without generating entanglement in the evolution~\cite{hovhannisyan2013entanglement}. Since then, a great deal of work has been devoted to understand the  performance of these devices under different scenarios. During the last years, general bounds for the power of collective processes where obtained~\cite{campaioli2017enhancing, julia2020bounds,gyhm2022quantum}, and also physical systems that were able to reach them were put forward~\cite{binder2015quantacell, ferraro2018high,crescente2020ultrafast}. In fact, it was recently shown that the \emph{collective advantage}~\cite{campaioli2017enhancing} for these batteries scales linearly with their number in the best case scenario~\cite{campaioli2017enhancing, julia2020bounds} and linearly with the order of the interaction in more realistic situations~\cite{gyhm2022quantum}. There were also several proposals for practical implementations of these devices using cavity QED systems~\cite{ferraro2018high, pirmoradian2019aging, andolina2019extractable}, many-body systems~\cite{binder2015quantacell, rosa2020ultra, ghosh2020enhancement, le2018spin,andolina2019quantum} and also batteries that interact with the environment~\cite{carrega2020dissipative,quach2020using, quach2022superabsorption}. 
The presence of quantum coherences is another feature whose role is expected to be important in the behavior of quantum batteries as well in other thermodynamic processes~\cite{binder2018thermodynamics, vinjanampathy2016quantum, goold2016role}.
In fact, recent studies show that quantum coherence may enhance the charging power of quantum batteries~\cite{monsel2020energetic,seah2021quantum, kamin2020entanglement} and the performance of some thermal machines~\cite{camati2019coherence, dann2020quantum,hammam2021optimizing, tajima2021superconducting, mitchison2015coherence,latune2019quantum,uzdin2016coherence,altintas2015rabi,rodrigues2019thermodynamics, rahav2012heat,son2021monitoring,streltsov2017colloquium,thingna2019landau}.

More recently, batteries described by open quantum systems, the so-called dissipative quantum batteries, were introduced~\cite{barra2019dissipative}. In this framework, the charging process is developed while the battery is coupled to a heat bath, and the dynamics is engineered in such a way that the steady state corresponds to a charged battery. In this context, also collective effects have been addressed  showing that the deterioration in the charging process due to dissipation could be mitigated by increasing the number of batteries~\cite{carrasco2021collective}.
This type of evolution has some advantages and drawbacks over the unitary one. On one hand, since it is a process that is designed so that the battery reaches an equilibrium state, one is not  concerned about its initial state and can always apply the same evolution. Moreover, since the steady state is of equilibrium~\cite{barra2017stochastic, barra2019dissipative}, keeping the charging process running while the battery is fully charged protects its state  against perturbations at no energy cost. The main drawback of this process, is that it dissipates some of the external work as heat, and thus it has a lower efficiency. 

In this paper, we consider different dissipative charging processes~\cite{barra2019dissipative} and study how collective effects and the presence quantum coherences affect their performance. 
The dissipative quantum batteries we considered are based on collisional models~\cite{barra2015thermodynamic, de2018reconciliation, karevski2009quantum, landi2014flux, de2020quantum, strasberg2017quantum, cattaneo2021collision,seah2019nonequilibrium, ciccarello2022quantum, campbell2021collision, manzano2022non, roman2021enhanced,hammam2022exploiting}, where the interaction between the battery and the bath is modelled in terms of a repeated interactions scheme~\cite{barra2019dissipative, de2018reconciliation} where many identical non-interacting systems in thermal equilibrium interact sequentially with the battery. This leads to a Markovian evolution that can be described in terms of a Lindblad equation~\cite{lindblad1976generators}. In this way, 
the process can be engineered in order to lead the battery towards non-equilibrium steady states that are not passive, i.e. work can be extracted from them. Here we show that within this framework the charging power of the collective process for $N$ batteries can scale as $N^3$ in the low temperature regime. Then, we show that also the presence of quantum coherences can enhance the charging power of a single battery, and finally we show how this process can be included into a finite time thermodynamic cycle.

The present work is structured as follows. In Sec.~\ref{sec:correlations} we study a collective dissipative battery charging process, focusing first on the relation between correlations and charging power. Then, we consider the power scaling law in terms of the number of subsystems conforming the battery. In Sec.~\ref{sec:coherehce_battery} we study a charging process in which coherences in the energy eigenbasis are generated by driving the battery with a Hamiltonian that does not commute with itself at all times. We find that coherence generation not only enhances the charging power, but also the efficiency of the process, that is the ergotropy over the external work. Then, we extend these results to a heat engine by introducing a cycle that includes the coherent dissipative charging process as one of its strokes. Finally, in Sec.~\ref{sec:conclusions} we exhibit the summary and conclusions.

\section{Collective dissipative charging}
\label{sec:correlations}
\begin{figure*}[tb]
    \centering
    \includegraphics[width = 0.7\textwidth]{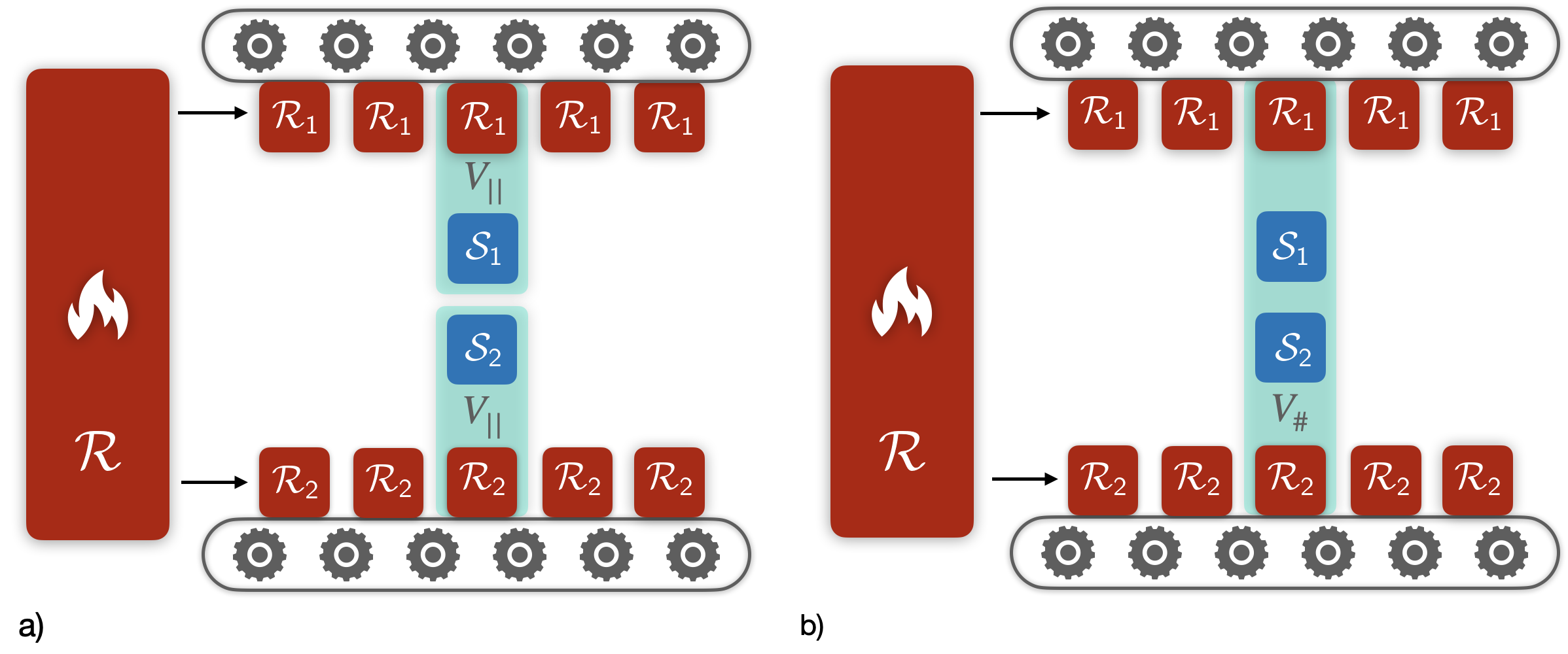}
    \caption{Dissipative charging process of two batteries, $\mathcal S_1$ and $\mathcal S_2$, for  parallel (a) and collective (b) schemes. The open dynamics is described in terms of a collision model. In this way, the batteries interact sequentially with auxiliary systems $\mathcal R_1$ and $\mathcal R_2$, that are prepared in thermal Gibbs states at the temperature of the heat bath $\mathcal R$, and discarded after the interaction. (a) In the parallel scheme each battery interacts with a single  auxiliary system at each time step. (b) In the collective scheme, we allow the batteries to interact with both auxiliary systems and with each other.}
    \label{fig:repeatedinteractions}
\end{figure*}
Collective processes can enhance the charging power of quantum batteries~\cite{hovhannisyan2013entanglement, campaioli2017enhancing, ferraro2018high}. For classical batteries, power scales linearly with the number of batteries, however, for quantum batteries this scaling can be superlinear. In fact, it has been shown that for unitary charging, the best possible scenario achieves a scaling that is quadratic~\cite{campaioli2017enhancing}.
In this work, we will be interested in the characterization of collective effects for dissipative charging processes~\cite{barra2019dissipative}.

We will consider the battery as a finite dimensional quantum system $\mathcal S$ with Hamiltonian $H_\mathcal{S}$ that is coupled to a heat bath $\mathcal{R}$. The open dynamics for the battery will be described in terms of a collision model~\cite{strasberg2017quantum, barra2015thermodynamic}. Within this approach the heat bath is composed by an ensemble of identical auxiliary systems, each with Hamiltonian $H_\mathcal{R}$, that are prepared in the same thermal Gibbs state at inverse temperature $\beta$. In this way, the evolution is such that each of these auxiliary systems interacts sequentially with the battery through a time-independent potential $V$ during a short time $\Delta t$.  Thus, at each time step the battery interacts with a different auxiliary system as it is schematically described in Fig.~\ref{fig:repeatedinteractions}. 
Notably, this model leads to a Markovian evolution for the battery whose steady state is of the form~\cite{barra2019dissipative}:
\begin{equation}
    \rho\,=\,\frac{e^{-\beta H_0}}{Z_0},
\end{equation}
in this case $H_0$ is an operator that satisfies:
\begin{equation}
    [H_\mathcal{S}, H_0]=0
    \quad
{\rm and}
\quad
  [V, H_0 + H_\mathcal{R}] = 0,
  \label{eq:commutation_relation}
\end{equation}
 and $Z_0=\Tr[e^{-\beta H_0}]$ ($\hbar=k_B=1$ throughout the entire work). 
 In general, $H_\mathcal{R}$ and $V$ do not determine uniquely $H_0$~\cite{barra2017stochastic}, meaning that the process may have more than one steady state. However, by a proper choice of $V$ and $H_\mathcal{R}$ it is possible to find a unique $H_0$ that satisfies the above commutation relation. Thus, in this case, the battery has a unique steady state that is independent of its initial state. Therefore, the process can be engineered, by a proper choice of these operators, so that asymptotically one ends up with a charged battery. Interestingly, in such a case if $H_0 = -H_\mathcal{S}$ the steady state of the process is thermal but with negative temperature. In addition, if the Hamiltonian spectrum is symmetric, this process maximizes the ergotropy~\cite{allahverdyan2004maximal}, since this state can be transformed via unitary operations into a thermal state~\cite{barra2019dissipative, alicki2013entanglement}. 
 
The type processes we described are such that the interaction between battery and the auxiliary systems does not commute with the total hamiltonian. Therefore, since the energy of the joint system is not preserved, some finite work is done to the system after each interaction. A fraction of this work goes to the battery while the rest goes to the bath as dissipated heat. Thus, the efficiency of the process can be defined as $\eta = 1 - Q/W$~\cite{barra2019dissipative}, where $W$ is the external work and $Q$ the dissipated heat.

In order to address the collective effects in dissipative charging processes, we will consider two different scenarios (see  Fig.~\ref{fig:repeatedinteractions}): a parallel one, where the batteries do not interact with each other, and a collective one in which the batteries interact with each other during the evolution. Each battery will be represented by a two-level system with Hamiltonian:
\begin{equation}
    H_\mathcal{S}\,=\,\frac{\omega}{2}\,\sigma_z.
\end{equation}
We will also consider auxiliary systems with Hamiltonian $H_\mathcal{R}\,=\,H_\mathcal{S}$ and choose $V$ in such a way that the steady state has a negative temperature.
   In this case, the efficiency of the charging process for a single battery is $\eta=1/2$~\cite{barra2019dissipative}, since half of the work done on the joint system is dissipated to the bath. Notice that if the auxiliary systems have a different gap, that is $H_\mathcal{R} = \alpha\,H_\mathcal{S}$, the steady state of the battery has a negative temperature  $ - (\beta \alpha)^{-1}$. That, using the expressions of work and heat for dissipative charging~\cite{barra2019dissipative}, leads to an efficiency $\eta\,=\,1/(1+\alpha)$. Therefore,  there is a trade-off between efficiency and ergotropy, meaning that when $\eta\rightarrow 1$ ($\alpha\rightarrow 0$) the ergotropy is vanishingly small.

\subsection{Collective charging of two batteries}

Let us first consider the simplest scenario with two batteries. 
As we said, we assume that all batteries have identical time independent hamiltonians, and the choice of charging interactions is what actually determines the nature of the process. 
For the parallel process we consider the following time independent potential~\cite{barra2019dissipative}:
\begin{equation}
        V_{\vert\vert}=\frac{\epsilon}{\sqrt{\Delta t}}\left(\sigma_{\mathcal{S}_1}^+\sigma_{\mathcal{R}_1}^+ +\sigma_{\mathcal{S}_2}^+\sigma_{\mathcal{R}_2}^+ \right)+  {\rm h.c.},
    \label{paralell_int}
\end{equation}
were $\epsilon$ is the coupling strength and $\Delta t$ is the time interval in which the system interacts with the auxiliary systems. 
The collective process is  designed  in such way that the commutation relation of Eq.~\eqref{eq:commutation_relation} is satisfied for $H_0=-H_\mathcal{S}$.
Thus, in this case we consider the following interaction:
\begin{equation}
       V_{\#}=\frac{2\epsilon}{\sqrt{\Delta t}}\left(\sigma_{\mathcal{S}_1}^+\sigma_{\mathcal{R}_1}^+\sigma_{\mathcal{S}_2}^+\sigma_{\mathcal{R}_2}^+ +\sigma_{\mathcal{S}_1}^+\sigma_{\mathcal{R}_1}^+\sigma_{\mathcal{S}_2}^-\sigma_{\mathcal{R}_2}^-\right)+ {\rm h.c.}
\label{collective_int}
\end{equation}
It can be shown that for each of these processes there is unique steady state that is thermal with a negative temperature whose magnitude equals the temperature of the bath. 
This is due to the fact that both interactions satisfy the commutation relations Eq.~\eqref{eq:commutation_relation} with $H_0=-H_\mathcal{S}$ as the only solution, which can be easily verified by replacing both interactions and the bath hamiltonian (see Eq.\eqref{eq:commutation_relation}). The factor 2 that appears in Eq.~\eqref{collective_int} is included to ensure that $\vert\vert V_\#\vert\vert_{op}\,=\,\vert\vert V_{\vert\vert}\vert\vert_{op}$. This condition was also imposed in~\cite{campaioli2017enhancing, binder2015quantacell, julia2020bounds}, in order to make a fair comparison between parallel and collective unitary processes. 
In this way, an optimal power scaling that is quadratic with the number of batteries was derived. When other condition is imposed, such as keeping constant the time-averaged energy for both processes, the same power scaling is obtained~\cite{campaioli2017enhancing}. However, this last condition is meaningless in our case, since we can make the collective process arbitrarily fast while keeping constant the time-averaged energy.

The process starts with empty batteries, which means that their initial state is thermal, and it ends when the batteries reach the asymptotic state.   In order to determine the equilibration time-scale for these processes, first we should notice that these collision models lead to Lindblad master equations in the limit of $\Delta t\rightarrow 0$~\cite{de2018reconciliation, barra2015thermodynamic}. A detailed derivation of the master equation in this limit can be found in Ref.~\cite{de2018reconciliation}. After solving these master equations, one finds that the batteries approach the steady state exponentially fast, with a characteristic time-scale $\tau$ that depends on the type of process:

\begin{equation}
    \rho(t)\,=\,\left[\,\rho_{empty}\,-\,\rho_{full}\,\right]e^{-t/\tau}\,+\,\rho_{full},
    \label{eq:rho(t)}
\end{equation}
were $\rho_{empty}$ is the initial thermal state, and $\rho_{full}=e^{\beta H_{\mathcal{S}_1}}/Z_1\otimes e^{\beta H_{\mathcal{S}_2}}/Z_1$ is the steady state, the two batteries in a negative temperature thermal state with $Z_1$ the partition function of a single battery.
Although for the collective process this time is temperature dependent, it is always less than half the time of the parallel process:
\begin{eqnarray}
    && \tau_{\vert\vert}\,=\,\frac{1}{\epsilon^2}, \nonumber \\
    && \tau_\#\,=\,\frac{1}{2\epsilon^2(1\,+\,\tanh^2(\beta \omega/2))},
    \label{eq:tau}
    \end{eqnarray}
 and is faster for lower temperatures. Thus, the collective process has a greater power than the parallel one.
 
Given that the mean power of a given process is   
\begin{equation*}
    P\,=\,\frac{E_{\delta}\,-\,E_{empty}}{T},
\end{equation*}
where $T$ is the time it takes to charge the battery with energy $E_\delta = E_{full}(1 - \delta)$, $\delta$ being a small number and $E_{full}$ the energy of the fully charged battery. We can now introduce the collective advantage~\cite{campaioli2017enhancing}: 
\begin{equation}
    \Gamma\,=\,\frac{P_\#}{P_{\vert\vert}}, 
\end{equation}
where $P_\#$ ($P_{\vert\vert}$) is the mean power of the collective (parallel) process. 

For two batteries, from Eq.~\eqref{eq:rho(t)}, we find that 
\begin{equation*}
    T\,=\,\tau\,\log\left[\frac{E_{full}\,-\,E_{empty}}{\delta\,E_{full}}\right].
\end{equation*}
As it can be seen, the only dependence on the type of process appears in $\tau$. Since both processes charge the battery to the same energy, the collective advantage $\Gamma$ only depends on the duration of each process. Thus, the collective advantage is:
\begin{equation}
    \Gamma\,=\,\frac{\tau_{\vert\vert}}{\tau_{\#}}\,=\,2\,\left(1\,+\,\tanh^2(\beta\omega/2)\right).
    \label{eq:advantage}
\end{equation}
Thus, for two batteries $2\leq\Gamma\leq 4$. 
In this simple example it is shown that the advantage could be grater than linear in the number of batteries, while previous works showed that $\Gamma\leq N$ for unitary quantum batteries~\cite{campaioli2017enhancing, julia2020bounds}.

Here we can also evaluate the relation between the collective advantage and the amount of correlations that are generated in the process. So we consider the mutual information~\cite{nielsen2000quantum} between the two batteries as a measure of total correlations. In particular, we compute the maximum mutual information along the process.
Using the fact that the evolution of the joint system is such that only the populations of the lowest and highest level are changed, and that the density matrix remains diagonal during the whole process, we can maximize the mutual information with respect to these two populations. In this way, we find that this maximum is reached when the populations of these two levels are equal. We can also notice that the populations of the intermediate levels are the corresponding to a thermal state of the joint system, and  the mutual information can be expressed as:
\begin{equation}
    \mathcal{I}_{max}\,=\,2\log 2\,+\,\left(\frac{Z\,-\,2}{Z}\right)\log\left(\frac{Z\,-\,2}{2Z}\right)\,-\frac{2}{Z}\log Z, 
    \nonumber
\end{equation}
where $Z= Z_1^2$. Now, using the temperature dependence of the collective advantage (Eq.\eqref{eq:advantage})  the mutual information can also be written as:
\begin{equation}
    \mathcal{I}_{max}\,=\,2\log 2\,+\,\frac{\Gamma}{4}\log\frac{\Gamma}{8}\,+\left(1-\frac{\Gamma}{4}\right)\log \left(\frac{1}{2}-\frac{\Gamma}{8}\right). \nonumber
\end{equation}
Since $2\leq\Gamma\leq 4$, we can see that the collective advantage grows monotonically with the maximal amount of correlations generated during the process. Thus, for two batteries we showed that there is an advantage in considering the collective process, and there is a direct relation between this advantage and the amount of correlations that are generated in the process. Finally, it is interesting to notice that the correlations generated in the process are classical. We calculated the quantum discord~\cite{ollivier2001quantum, adesso2016introduction} and it turns out to be zero for this interaction. So this enhancement in power does not rely on the generation of quantum correlations. 
In the following, we will generalize these ideas to an arbitrary number of batteries.

\subsection{Scaling of the charging power}
We are interested in evaluate the scaling of the charging power with the number of batteries $N$. Thus, in order to include more batteries we have to design the  interaction that leads the system to the proper steady state. This implies that the interaction should satisfy $[V_\#^{(N)}, -H^{(N)}_\mathcal{S} + H^{(N)}_A] = 0$, where $H^{(N)}$ is the sum Hamiltonian of the $N$ systems. 

We find that the interaction that generates the desired process is the one that exchanges the populations between the energy levels $E_k\longleftrightarrow E_{N-k}$.  Thus, the operator that performs this action over the energy levels $j$ and $l$ of the batteries and the bath is: 
 $\hat V_{j,l}\equiv \sum_{\mu_j,\mu_l} \ketbra{E_{j{\mu_j}}, E^{(\mathcal R)}_{j{\mu_j}}}{E
    _{l\mu_l},E^{(\mathcal R)}_{l\mu_l}}$, 
    where $\mu_j=1,...,g_j$ and ${g_j=\binom{N}{j}}$ is the degeneracy of the {$j$-th} energy level.  In this way, the collective potential can be written as:

\begin{equation*}
    V_\#^{(N)}=\sum_{k=1}^N\frac{N\epsilon}{\sqrt{\Delta t}} \, \hat V_{k,N-k},
    \end{equation*}
It is important to remark here that since we require that the parallel and collective interactions have the same operator norm, the coupling constant must scale as $\epsilon\rightarrow N\epsilon$. This scaling is what leads to a speed-up of the process.
Using this interaction, and following the same steps as in Ref.~\cite{de2018reconciliation}, we can obtain the following master equation: 
\begin{equation*}
    \Dot{\rho_\mathcal{S}}\,=\,-i\comm{H_\mathcal{S}^{(N)}}{ \rho_\mathcal{S}}\,+\,D(\rho_\mathcal{S}),
\end{equation*}
where 
\begin{equation*}
    D(\rho_\mathcal{S})\,=\,-\frac{1}{2}\,\Tr_\mathcal{R}\,\comm{V_\#^{(N)}}{\comm{V_\#^{(N)}}{\rho_\mathcal{S}\otimes\,\frac{e^{-\beta\,H_\mathcal{R}^{(N)}}}{Z_1^N}}}.
\end{equation*}
From these equations it can be shown that
the system has a negative temperature thermal state as the only steady state. 
Then, by solving the master equation we find that, for an initial thermal state, the energy populations approach the steady values exponentially in time. However, the characteristic time-scale is different for each energy level. In fact, for $N$ batteries the mean value of the energy evolves as:
\begin{equation}
    E(t)=\sum_k g_k\left[\frac{\left(e^{-\beta E_k}-e^{\beta E_k}\right)}{Z_1^N}e^{-t/\tau_k}+\frac{e^{\beta E_k}}{Z_1^N}\right]E_k, \nonumber
\end{equation}
where 
\begin{equation}
    \tau_k\,=\,\frac{Z_1^N}{2\cosh{(\beta E_k)}N^2\epsilon^2}
\end{equation}
is the characteristic time-scale for the transition ${E_k\longleftrightarrow E_{N-k}}$.
It is interesting to notice that these characteristic time-scales are in general greater than the ones obtained in the parallel process, 
 except for the transition between the highest and lowest levels. This means that in general, the collective process is slower than the parallel one. However, in the low temperature regime $k_BT\ll \hbar\omega$, the only relevant transitions are those that exchange the populations between the lowest and the highest energy levels. 
Therefore, in this regime, we find that the collective process is faster than the parallel one. More importantly, when studying the scaling of the collective advantage we find that 
\beq
\Gamma\leq N^2,
\eeq
where the upper bound is attained at zero temperature.
Notably, this upper bound is $N$ times greater than the one obtained for unitary batteries~\cite{campaioli2017enhancing,julia2020bounds}.
This is due to the fact that in our case the quantum advantage (in the low temperature regime) scales quadratically with the coupling constant  while for instance in~\cite{julia2020bounds} it does linearly. 

\begin{figure}[tb]
    \centering
    \includegraphics[width = 8.6cm,height = 70mm]{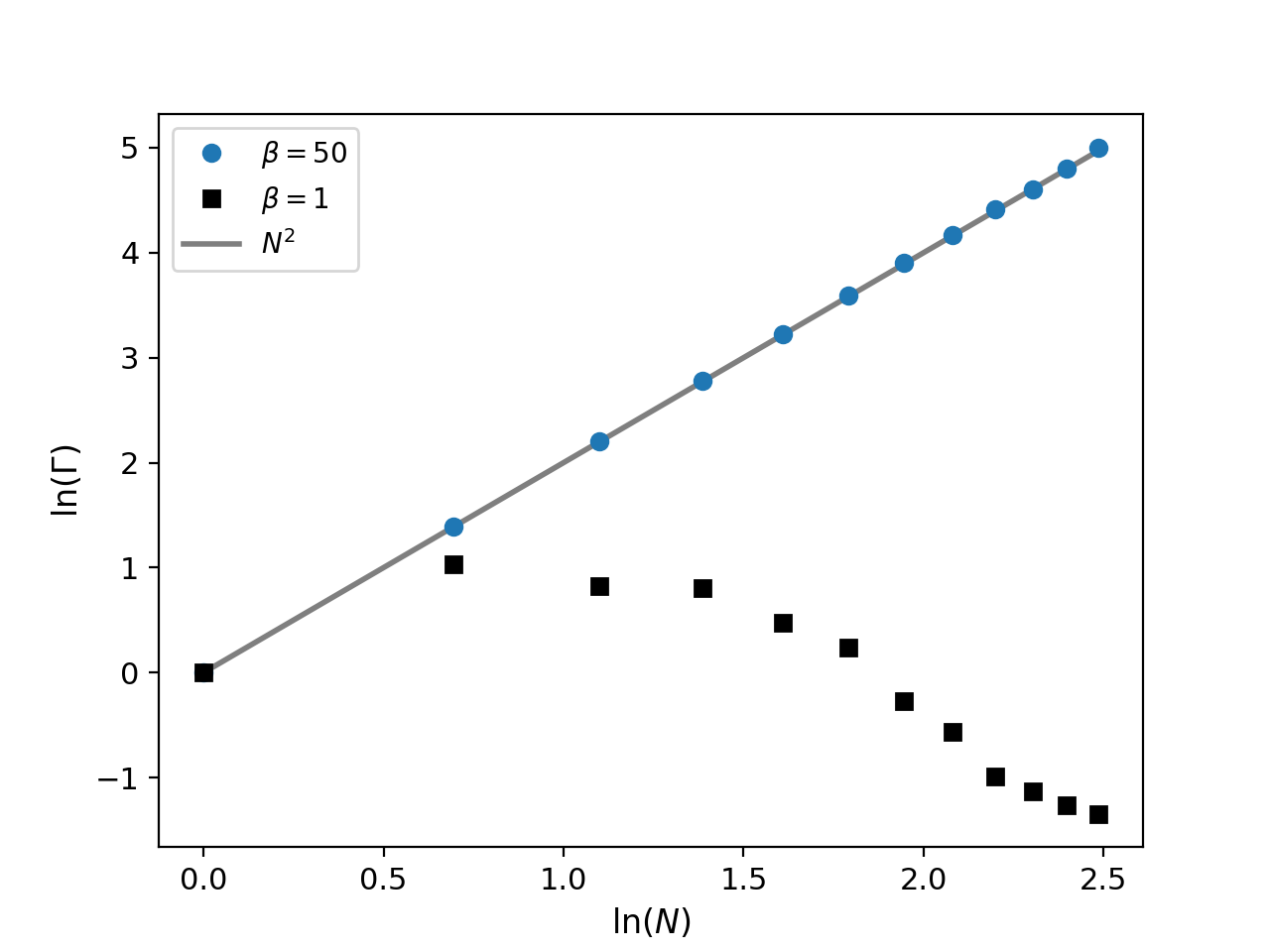}
    \caption{Collective advantage vs. the number of batteries $N$ in log scale for two different temperatures. $\ln(N^2)$ is plotted to show that it fits the scaling of the collective advantage for the lower temperature (gray line). At low temperatures (blue dots) the collective advantage is quadratic, while at higher temperatures (black squares) the collective process we designed has less power than the parallel one. The simulations were done using the following parameters: $\epsilon=\sqrt{10}$, $\omega=1.5$, $\delta=0.01$. }
    \label{fig:P_vs_N}
\end{figure}

\begin{figure}[tb]
    \centering
    \includegraphics[width = 8.6cm,height = 70mm]{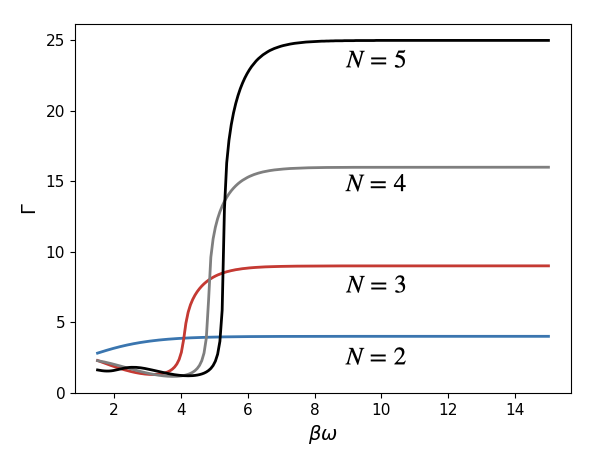}
    \caption{Collective advantage vs. inverse temperature for $N$ batteries. $\Gamma$ reaches the quadratic regime at low temperatures. As $N$ increases, this is achieved at a higher value of $\beta\omega$. Parameters: $\delta=0.01$.}
    \label{fig:Gamma_betas}
\end{figure}

We have performed numerical simulations to verify this scaling,
see Fig.~\ref{fig:P_vs_N}. There we can see that for low temperatures $\Gamma$ is quadratic with the number of batteries, while for high temperatures this interaction leads to a process with worse power than the parallel one.
In Fig.~\ref{fig:Gamma_betas}, we show the collective advantage in terms of $\beta\omega$ for up to five batteries. There we can see that in all the cases $\Gamma\rightarrow N^2$ when ${\beta\omega\gg 1}$, and this regime is achieved for temperatures of the order of $\beta\omega\approx 10$.
It is important to remark that, even tough this process allows for a greater collective advantage when compared to the closed system scenario, it is less efficient, in the sense that some of the external work done to charge the batteries is dissipated as heat.

We showed that when dealing with dissipative batteries, the power scaling is not upper bounded by $N^2$. We have obtained this behavior for a specific choice of the interaction $V_{\#}$ that includes ${N-body}$ interactions. However, one can also notice that if one allows only ${k-body}$ interactions one can charge the batteries in partitions of $k$ batteries. In this case, each set has a collective advantage of $\sim k^2$ in the low temperature regime, thus
the collective advantage of the $N$ batteries  is at least also $\Gamma\sim k^2$,
while for unitary batteries $\Gamma \sim k$~\cite{gyhm2022quantum}.

\section{Dissipative charging process with coherence generation}

\label{sec:coherehce_battery}
In this section we will study another feature that may affect the charging process:  the generation of coherences during the dynamics.
The dissipative charging process we considered in the previous section was unable to generate coherences in the energy basis. This is due to the fact that the dissipator does not couple the off-diagonal and diagonal terms of the battery's density matrix, so it commutes with the Hamiltonian at all times. 
In order to generate coherences in the dynamics, we will introduce a time dependent Hamiltonian that does not commute with itself at all times: 
\begin{equation}
    H_\mathcal{S}(t)=\frac{\omega}{2}\left[\alpha(t)\,\sigma_x+(1-\alpha(t))\,\sigma_z\right],\,\alpha(t)\,\in\,[0,1], 
    \label{eq:Hamilttime}
\end{equation}
where $\alpha(t)$ is a real time-dependent parameter, and for $\alpha(t)=0$ is equivalent to the previous process. 
Whereas the interaction that couples the battery to the bath is the same that we used for single batteries in the previous Section:
\begin{equation}
    V\,=\,\frac{\epsilon}{\sqrt{\Delta t}}\left(\sigma_\mathcal{S}^+\sigma_\mathcal{R}^+\,+\sigma_\mathcal{S}^-\sigma_\mathcal{R}^-\right).
    \label{eq:interactionSingle}
\end{equation}
The other condition we require is that $\alpha(t)$ returns to zero at some point so as to reach the desired steady state, otherwise the commutation relation that defines $H_0$ would not be satisfied.

\subsection{Charging power of a single dissipative battery}

In particular, we will analyze the process that maximizes power subjected to the time dependent Hamiltonian of Eq.~\eqref{eq:Hamilttime}. To this end, we use quantum control techniques~\cite{larocca2018quantum,khaneja2005optimal,PhysRevA.37.4950} to optimize power over $\alpha(t)$. 

The procedure that carries out this optimization goes as follows. First, we perform a time discretization by transforming the continuous parameter $\alpha(t)$ into a vector of parameters $\{\alpha_k\}$. This discretization simplifies the dynamics, since in this way it can be implemented in terms of a concatenation of short evolutions within time intervals $\Delta t_k\,=\,t_{k+1}-t_k$ and constant hamiltonians $\{H(\alpha_k)\}$:
\begin{equation}
    \rho_N\,=\,\Pi_{k=1}^N\,\exp\{\mathcal{M}(\alpha_k)\}\,\rho_0,
\end{equation}
where $\mathcal{M}(\alpha_k)$ is the propagator given by the master equation. Now, if we write $\rho$ as a column vector:
\begin{equation*}
    \rho\,=\,
    \begin{pmatrix}\rho_{00}\\
\rho_{01}\\
\rho_{10}\\
\rho_{11}
\end{pmatrix}
\end{equation*}
then
\begin{multline*}
\mathcal{M}(\alpha_k)=\\\begin{pmatrix}-\frac{\gamma^{+}}{2} & \frac{i\omega\alpha_k}{2} & -\frac{i\omega\alpha_k}{2} & \frac{\gamma^{-}}{2}\\
\frac{i\omega\alpha_k}{2} & -\frac{\epsilon^{2}}{4}-i\omega(1-\alpha_k) & 0 & -\frac{i\omega\alpha_k}{2}\\
-\frac{i\omega\alpha_k}{2} & 0 & -\frac{\epsilon^{2}}{4}+i\omega(1-\alpha_k) & \frac{i\omega\alpha_k}{2}\\
\frac{\gamma^{+}}{2} & -\frac{i\omega\alpha_k}{2} & \frac{i\omega\alpha_k}{2} & -\frac{\gamma^{-}}{2}.
\end{pmatrix}
\end{multline*}
where $\gamma^\pm = \epsilon^2\exp({\pm\beta\omega/2})/Z_1$ are the dissipation rates of the master equation.

For a two-level battery, the energy maximization is carried out through a maximization of excited state population. Then, by using gradient ascent method,  the parameters $\alpha_k$ can be updated from an arbitrary initial choice so as to maximize this population at any time $t_N$. This is done using the following procedure: First, we set $\alpha_k$ as a vector of random components; and then, $\alpha_k$ is updated with the following formula:
\begin{equation}
    \alpha_{k}^{(i+1)}\,=\,\alpha_k^{(i)}\,+\,\zeta\,\frac{\partial \rho_e(t_N)}{\partial\alpha_k}
    \nonumber
\end{equation}
where $\rho_e$ are the populations of the excited state and $\zeta$ is a parameter that determines the size of variation of $\alpha_k$ on each step. Using this procedure $\alpha_k$ is updated until it reaches the optimal parameters, for which the gradients become small enough that make the variations in $\alpha_k$ negligible.

Using this algorithm for different times $t_N$, we found that the process that maximizes power consists of two sudden quenches: one at the beginning of the process switching $\alpha$ from 0 to 1, and the other after some time $t_d$ (which results from the optimization and depends on $\epsilon$ and $\omega$) that returns $\alpha$ to 0. We have also found that the process with minimum power is the one that takes place at constant Hamiltonian proportional to $\sigma_z$. Notably, this is the only process of the family defined in Eq.~\eqref{eq:Hamilttime} that does not generate quantum coherence in the energy basis. 

\begin{figure}
    \centering
    \includegraphics[width = 8.6cm]{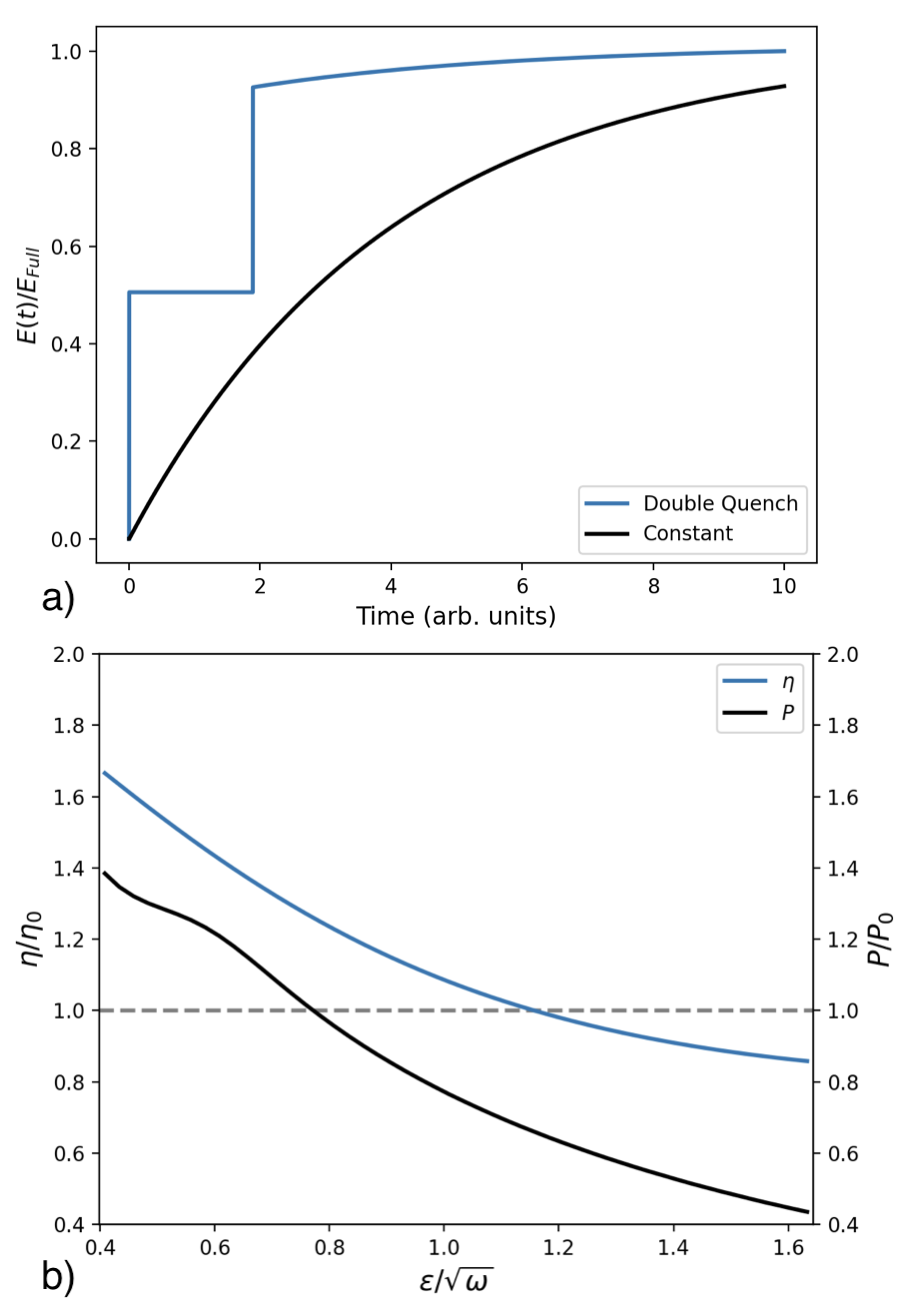}
    \caption{(a) Fraction of energy stored in the battery as a function of the duration of the charging process. (b) Efficiency (Power) of the coherent process normalized by the efficiency (power) of the constant Hamiltonian process, $\eta_0$ ($P_0$), plotted against the coupling strength. Parameters: $\epsilon=\sqrt{0.25}$, $\omega=1.5$, $t_d=2$, $\beta=1$.}
    \label{fig:doble_quench}
\end{figure}

In Fig.~\ref{fig:doble_quench}(a) we show the optimal process along with the standard dissipative process that does not generate coherences. It can be seen that the charging process is faster when using the double quench protocol. The plot is shown using instantaneous quenches, that is why we have infinite instant power for some times, however this enhancement due to the presence of coherences remains even for finite-time quenches. 
In Fig. \ref{fig:doble_quench}(b) we show the efficiency of the process $\eta$, i.e. the ergotropy over external work~\cite{barra2019dissipative}, and the power as a function of the coupling strength between the system and the reservoir. We find that in the weak coupling regime the double quench process is also more efficient than the process with constant Hamiltonian. This is due to the fact that for strong couplings the dissipative term dominates the dynamics. In fact, the advantage appears only when the system is initially in a passive state. On the other hand, when the system starts in an active state the driving slows down the charging process and produces more dissipation. For strong couplings the state remains passive just for a short time, and that is why we observe a reduction in efficiency and power as the coupling is increased.

\begin{figure}[tb]
    \centering
    \includegraphics[width = 8.6cm,height = 60mm]{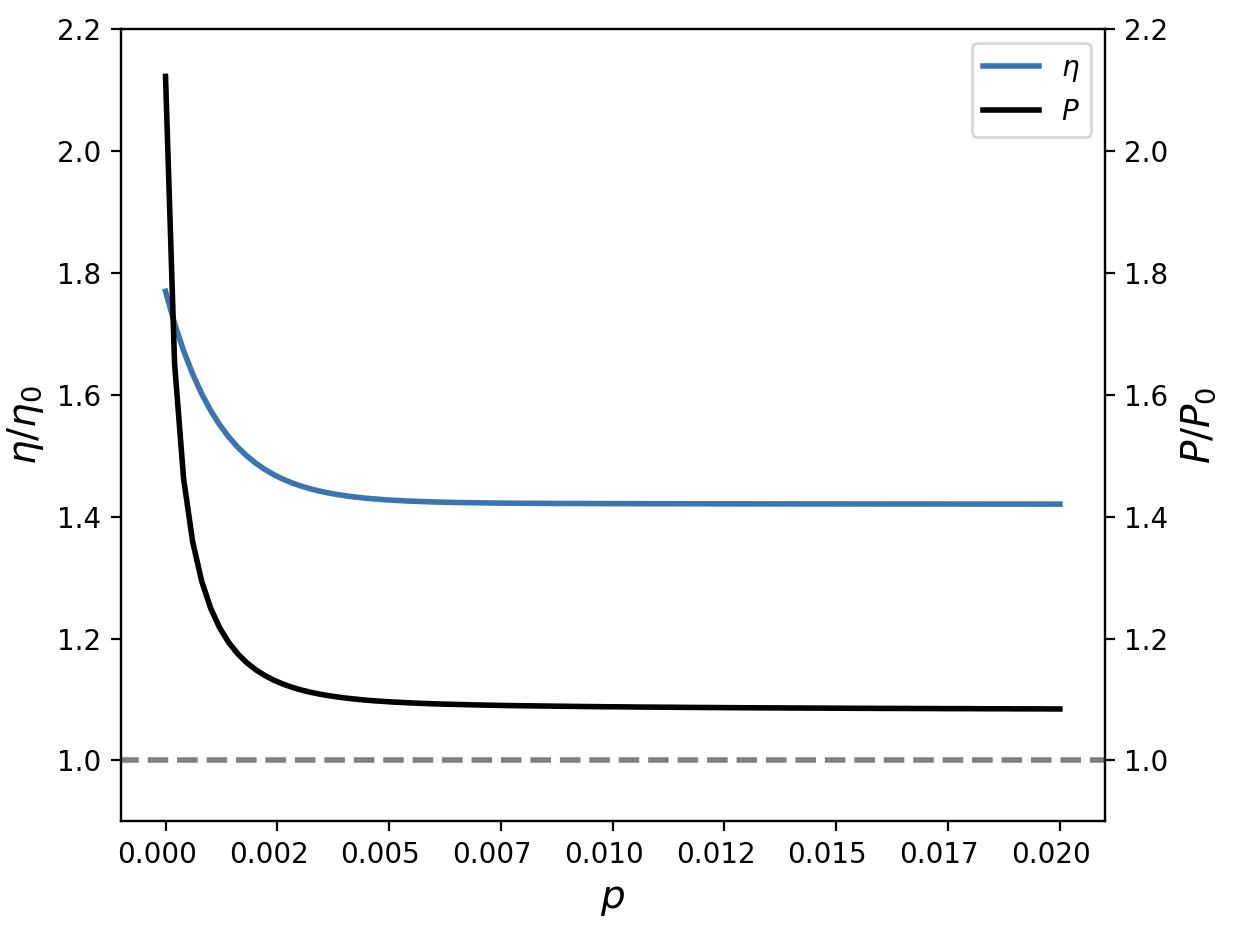}
    \caption{Power and efficiency of the coherent charging process plotted against the dephasing parameter $p$. Power (Efficiency) is normalized by the power (efficiency) of the constant process. Parameters: $\epsilon=\sqrt{0.25}$, $\omega=1.5$, $t_d=2$, $\beta=1$.}
    \label{fig:dephasing}
\end{figure}

In order to relate the enhancement in power and efficiency with the presence of coherences, we modify the process by introducing a \emph{dephasing noise} that is turned on during the whole evolution. This noise removes coherences in the energy eigenbasis with probability $p$:
\begin{equation}
    \mathcal{E}(\rho)\,=\,\left(1\,-\,\frac{p}{2}\right)\,\rho\,+\,\frac{p}{2}\,\sum_i\,\Pi_{E_i}\,\rho\,\Pi_{E_i}
    \label{eq:dephasing_noise}
\end{equation}
where $\Pi_{E_i}$ are projectors onto the instantaneous energy eigenstates. Thus, this noise does not change the energy of the system. In Fig.~\ref{fig:dephasing}, we show both power and efficiency in terms of the probability $p$. It can be seen that both magnitudes are quickly deteriorated as coherences are removed, so in this way the best performance of the process can be connected with the generation of quantum coherence in the energy eigenbasis. 

We should also notice that when coherences are completely removed, the process is still more efficient than the one with constant Hamiltonian. This happens because in a charging process with a time-dependent Hamiltonian, the state of the battery does not commute with the Hamiltonian at all times, therefore part of the work done on the battery comes from this time-dependence and it does not dissipate heat to the bath. On the other hand, when the Hamiltonian is constant, all the work comes from the interaction with the bath and more heat is dissipated to the environment. 

\subsection{Heat engine with coherence generation}
\label{sec:coherence_engine}
In the last section, we showed that the best performance of our charging process was associated with the presence of coherences. Thus, one might wonder whether this behavior could also be observed in a thermodynamic cycle.
 In fact, one of the most interesting questions in quantum thermodynamics is whether or not quantum coherence can enhance the performance of heat engines. Recent works~\cite{camati2019coherence, dann2020quantum} show that, under certain conditions, the generation of quantum coherence in the energy basis can enhance the efficiency and power of heat engines. Thus, in this section we study a thermodynamic cycle that includes the dissipative charging process we introduced as one of its strokes.
 
 The cycle we design is similar to the Otto cycle~\cite{feldmann1996heat}, see Fig.~\ref{fig:CicloActivo}. However, since we want to include the battery charging scheme in one of the strokes, we considered an Otto cycle in which the interaction between system and baths is the same as in Eq.~\eqref{eq:interactionSingle}.
In this way, the steady state of the open strokes will be at negative temperature:  
\begin{equation*}
\rho_{-\beta_i}(H_i)\,=\,\frac{e^{\beta_i H_i}}{Z_i}.\quad i=c,h.    
\end{equation*}
Where the index $i$ refers to the cold ($c$) and hot ($h$) baths. 
 In addition, since we found that the generation of coherence improves the charging process only when the system starts in a passive state, we include in one of the closed strokes a unitary transformation that inverts the populations of the system, leaving it in a thermal state. In this way, there is an open stroke where the system starts in a passive state and we can use the generation of coherences to improve their performance.  Bellow, we describe the thermodynamic cycle:
 
\begin{figure}[tb]
    \centering
    \includegraphics[width = 8.6cm,height = 50mm]{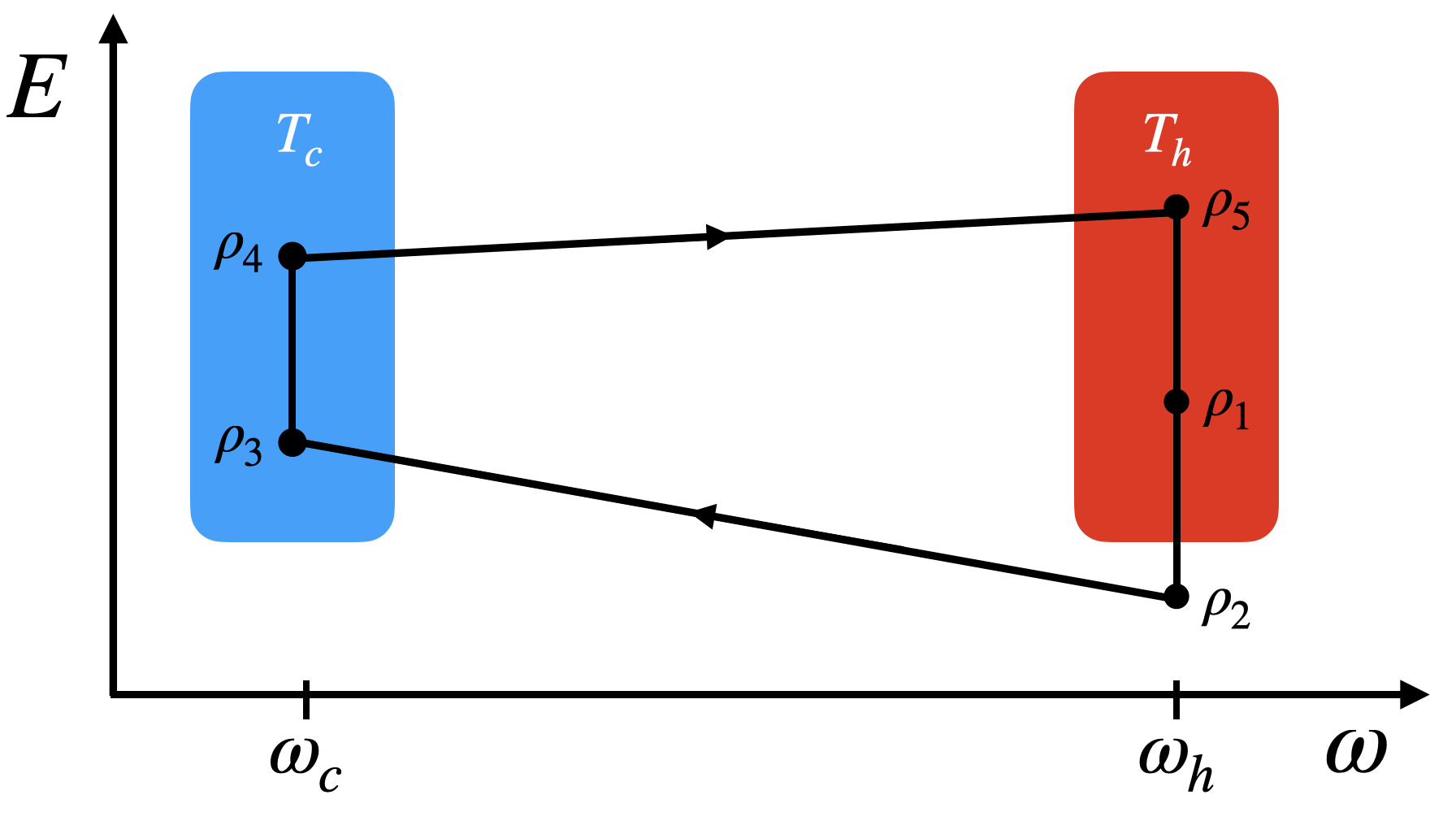}
    \caption{Schematic representation of the thermodynamic cycle. The \textit{y}-axis displays the energy of the system at each stroke, and the \textit{x}-axis its energy gap.
    }
    \label{fig:CicloActivo}
\end{figure}

\begin{enumerate}

    \item The system with Hamiltonian $H_h\,=\,\frac{\omega_h}{2}\sigma_z$ starts in a negative temperature state at the temperature of the hot bath $\rho_1\,=\,\rho_{-\beta_h}(H_h)$.  A unitary transformation that inverts the qubit's populations is performed, and the system ends up in the thermal state $\rho_2\,=\,\rho_{\beta_h}(H_h)$. During this stroke, the work extracted from the system is equal to:
    \begin{equation*}
        W_1\,=\,\Tr\left[H_h\,\left(\rho_2\,-\,\rho_1\right)\right].
    \end{equation*}
    
    \item The system is isolated from the baths, and undergoes an adiabatic compression by changing its gap from $\omega_h$ to $\omega_c$. Its state does not change, but  it performs negative work (external work is required):
    \begin{equation*}
        W_2\,=\,\Tr\left[\left(H_c\,-\,H_h\right)\,\rho_2\right]
    \end{equation*}
    
    \item This stroke consists of the dissipative charging of a quantum battery (the system) studied in the previous section. Thus, it is coupled to the cold reservoir and the double quench Hamiltonian is used to inject work generating coherences in the energy basis. In this case, the system starts in the state $\rho_3\,=\,\rho_2$ and ends in a negative temperature thermal state at the temperature of the cold reservoir $\rho_4\,=\,\rho_{-\beta_c}(H_c)$. During this stroke, some external work needs to be done, and some heat is dissipated to the bath. Analytical expressions for the heat and work are rather complicated and therefore omitted in the main text (see Appendix~\ref{sec:app}). 

    \item  The system is separated from the bath and the gap is changed to its initial value $\omega_c\rightarrow \omega_h$. During this stroke, the state remains constant $\rho_5\,=\,\rho_4$ and again some external work needs to be done:
    \begin{equation*}
        W_4\,=\,\Tr\left[\left(H_h\,-\,H_c\right)\,\rho_4\right].
    \end{equation*}
    
    \item Finally, to close the cycle, the system is coupled to the hot bath and it reaches the non-equilibrium steady state $\rho_1$. During this process, the  state absorbs heat:
    \begin{equation*}
        Q_h\,=\,\Tr[H_h\,\left(\rho_5\,-\,\rho_1\right)],
    \end{equation*}
    and performs some work:
    \begin{equation*}
        W_5\,=\,2\,\Tr[H_h\,\left(\rho_5\,-\,\rho_1\right)].
    \end{equation*}

\end{enumerate}

\begin{figure}[tb]
    \centering
    \includegraphics[width = 7.6cm]{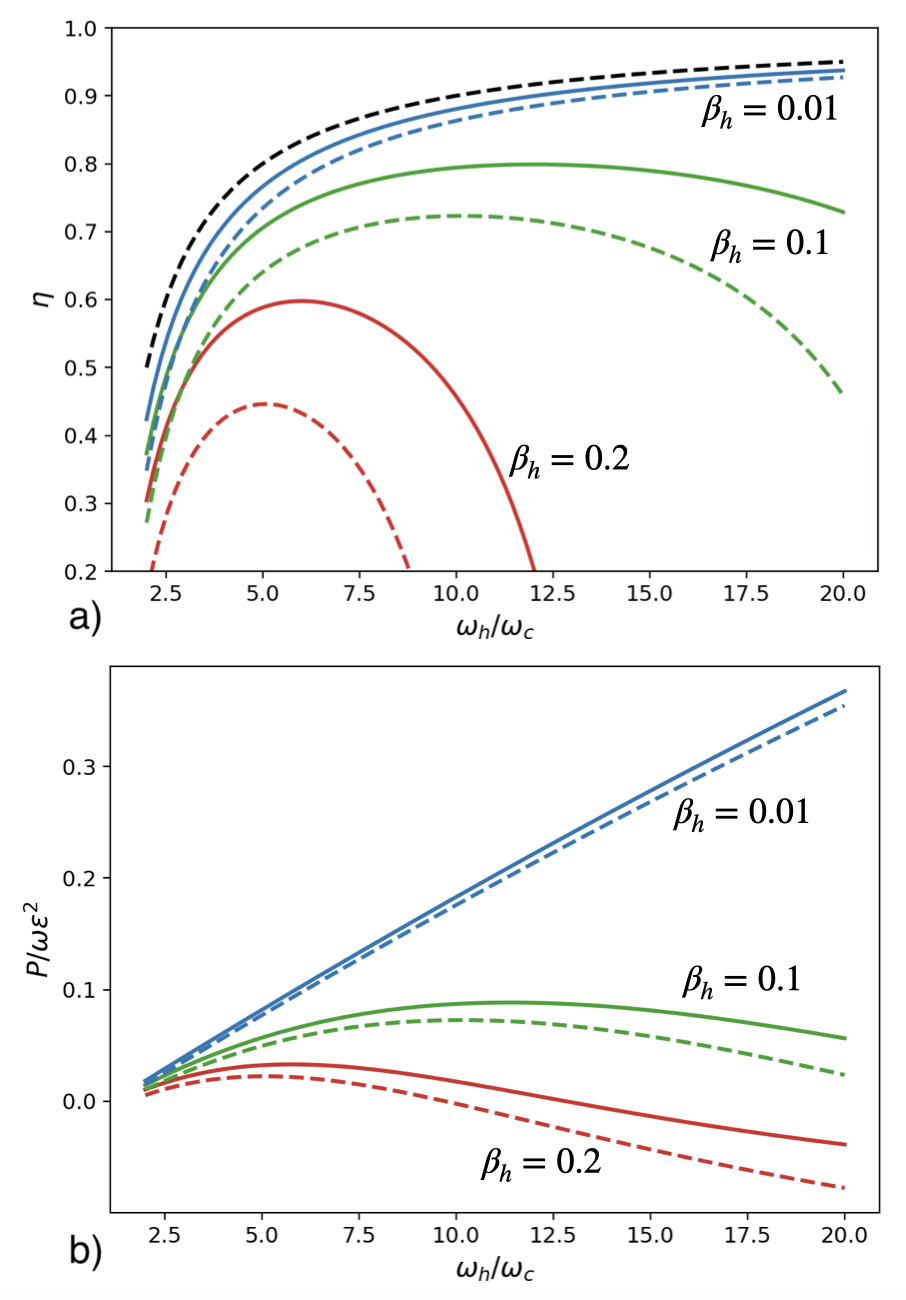}
    \caption{(a) Efficiency and (b) power of the coherent (solid lines) and dephased  (dashed lines) cycles for different temperatures of the hot bath in terms of systems' gap $\omega_{h}$. The black dashed line in (a) shows the equivalent Otto efficiency.
    Parameters: $\epsilon=\sqrt{0.25}$, $\omega_c=1$, $t_d=2$, $\beta_c=10$, $t_{cycle}=20$.
    }
    \label{fig:Eff_vs_T_vs_h}
\end{figure}

The efficiency of the process can be computed analytically and is included in the Appendix~\ref{sec:app}.
We performed numerical simulations to study the efficiency and power of this cycle operating at finite time for different parameters. The simulations where done both for a coherent cycle
and an incoherent one, in which coherences are eliminated using dephasing noise Eq.~\eqref{eq:dephasing_noise} with $p=1$. In this case, we consider cycles whose duration is larger than the relaxation time of the open strokes ($t_{cycle}\gg\epsilon^{-2}$). The results for efficiency at different temperatures $\beta_h$ and gaps $h_h$  are shown in Fig.~\ref{fig:Eff_vs_T_vs_h}(a). 
There we can see that the coherent cycle has a greater efficiency for all the parameters we considered. The same happens with power, 
as is shown in Fig.~\ref{fig:Eff_vs_T_vs_h}(b). 

\begin{figure}[tb]
    \centering
    \includegraphics[width = 8.6cm,height = 60mm]{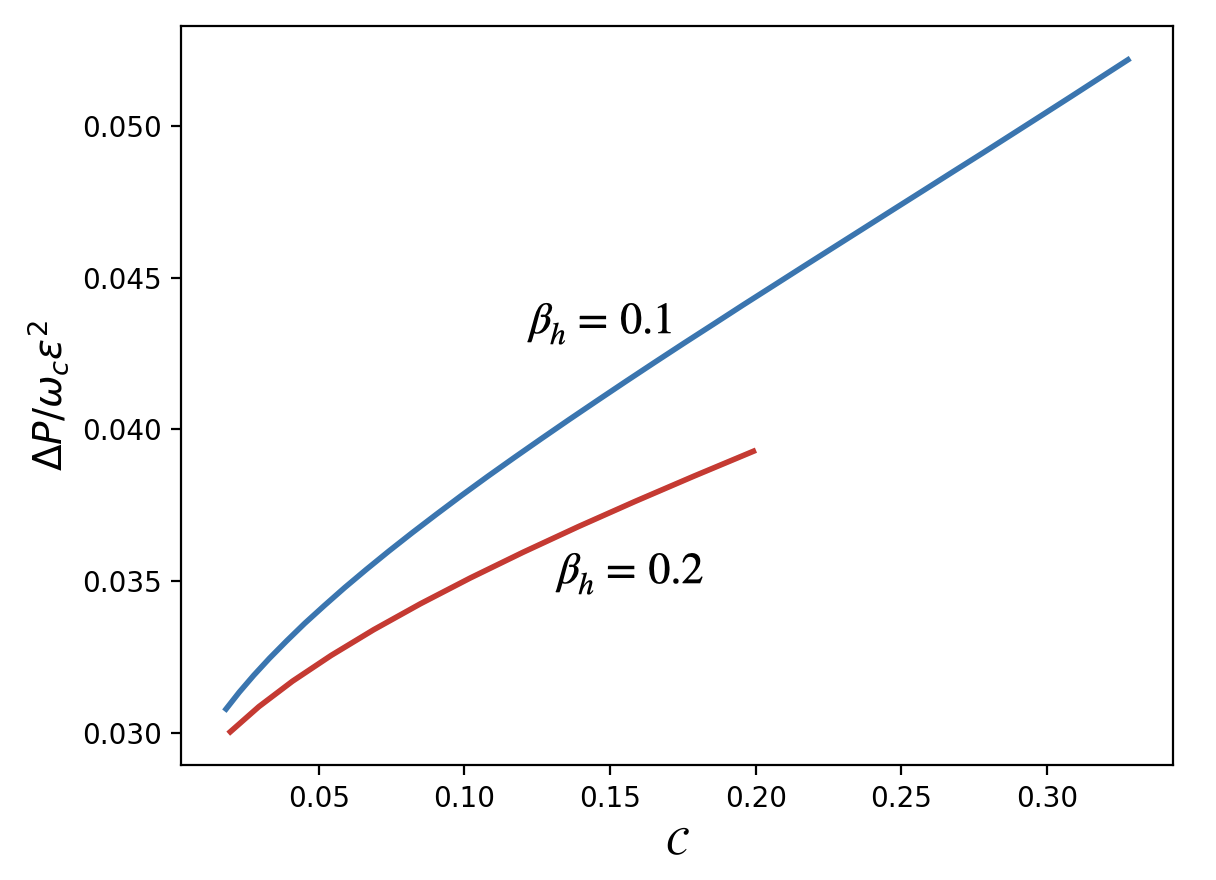}
    \caption{Difference in power between the coherent and dephased cycle in terms of the relative entropy of coherence. The enhancement in power grows monotonically with the maximum amount of coherence generated during the cycle. 
    The parameters are the same of Fig. \ref{fig:Eff_vs_T_vs_h}.}
    \label{fig:coherence}
\end{figure}

Here we show an example where the generation of coherence in the energy basis results in an enhancement of the cycle performance. Usually, quantum coherence reduces the efficiency of thermodynamic cycles due to an increased dissipation, this is called quantum friction~\cite{rezek2010reflections, PhysRevE.65.055102, PhysRevE.68.016101, PhysRevE.73.025107}, but in this case we find a process in which coherence has the effect of quantum lubrication~\cite{tajima2021superconducting, camati2019coherence}. However, both power and efficiency are always lower than those of an equivalent Otto engine operating between the same temperatures and gaps. 
We also evaluate the difference in power between the coherent and the dephased cycle as a function of the amount of coherence generated in the process. This was done using the relative entropy of coherence as a figure of merit:
\begin{equation}
    \mathcal{C}\,=\,\max_{\rho}\left\{S(\rho_{deph})\,-\,S(\rho)\right\}
\end{equation}
where $S(\cdot)$ is the von Neumann entropy and $\rho_{deph}=\sum_i\,\Pi_{E_i}\rho\Pi_{E_i}$ is the  dephased state. This quantity measures the maximum amount of coherence generated along the process. In Fig.~\ref{fig:coherence} we show that, as we expect, this power difference increases with the amount of coherence.

\begin{figure}[tb]
    \centering
    \includegraphics[width = 7.6cm]{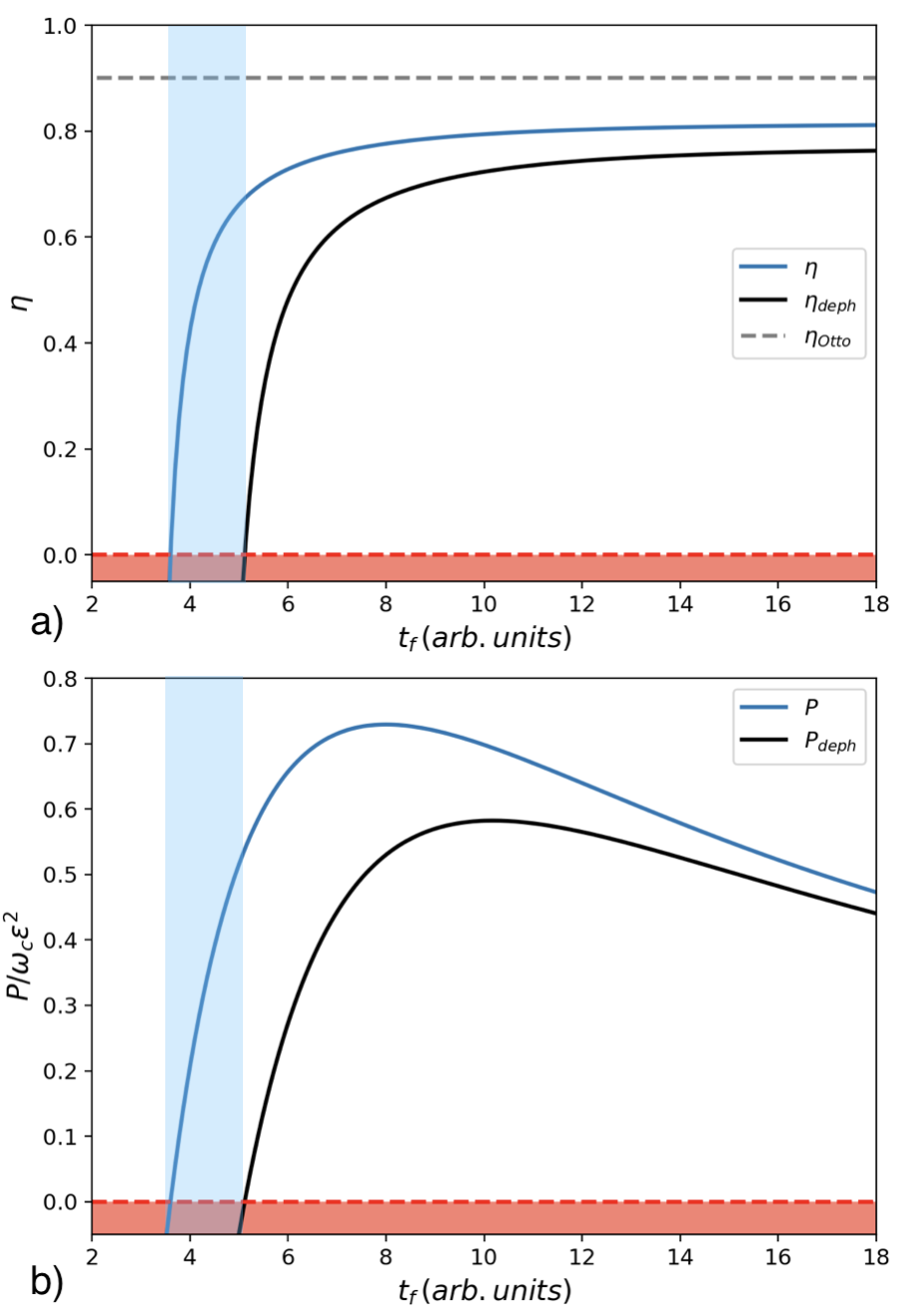}    
    \caption{(a) Efficiency and (b) power for the coherent (blue [light gray]) and dephased (black) cycles in terms of the duration of the cycle. In Fig. (a) the grey dashed line shows the equivalent Otto efficiency. The red area in the bottom of both figures highlights the region for which the cycles do not produce work. The light blue, vertical, shadowed areas show the range of cycle durations for which the cycle works as an engine only if coherences are generated.
    Parameters: $\epsilon=\sqrt{0.25}$, $\omega=1.5$, $t_d=2$, $\beta_c=10$, $\beta_h=0.1$.}
    \label{fig:finite_time}
\end{figure}

In previous works~\cite{camati2019coherence, dann2020quantum} short time thermodynamic cycles were also studied, showing that coherence generation can improve their performance even when the duration of the open strokes were much shorter than the equilibration time. Here we find the same behaviour for this sorts of cycles where $t_{cycle}\ll\epsilon^{-2}$. In Fig.~\ref{fig:finite_time}(a) we show the efficiency of the cycle as a function of its duration. 
There we can see that the coherent cycle has a greater efficiency for all times and it remains positive for short times where the incoherent cycle cannot produce work. A similar behaviour is observed in the power (Fig.~\ref{fig:finite_time}(b)). Thus,  the generation of coherences not only improves the efficiency and power of stationary long time cycles, but  also allows to extract work in short time cycles. 

\section{Summary and Conclusions}
\label{sec:conclusions}
In this work we have studied two important features of dissipative quantum batteries: their performance under collective charging and quantum coherences.
To this end, we design a collective charging process that involves interactions between the batteries, and compared it with a parallel process. For two batteries we have found that the collective process is faster than the parallel one. Both power and the amount of correlations depend on the temperature of the reservoir, and there is a direct relation between the amount of correlations and the power of the process.
Then, we investigate how this collective enhancement in power scales with the number of batteries $N$. We showed that in the low temperature regime the collective advantage scales as $\Gamma\propto N^2$ in our scheme. Notably, this scaling is grater than the one obtained for unitary charging processes, however it has lower efficiency. On the other hand, if one considers $2k-body$ interaction in the Hamiltonian, we showed that the collective advantage turns to be $\Gamma= k^2$.

When the charging process generates coherence in the energy basis, we have found that there is an improvement both in power and efficiency. We also showed that this advantage is quickly deteriorated when a dephasing noise that eliminates coherences is added. Finally, we studied a thermodynamic cycle that contains this charging process as one of its strokes. This cycle operates between non-equilibrium steady states, and displays a better performance when quantum coherences are developed in the evolution.

\acknowledgments
  The authors acknowledge support from CONICET, UBACyT and ANPCyT.

\appendix
\section{Efficiency of the cycle in the Weak coupling regime}
\label{sec:app}
In this appendix we derive the efficiency of the coherent cycle in the weak coupling regime, and show that it is always upper bounded by Otto efficiency.

The efficiency is defined as:
\begin{equation}
    \eta\,=\,\frac{W}{\vert Q_h\vert}\,=\,\frac{\vert Q_h\vert - \vert Q_c\vert}{\vert Q_h\vert}\,=\,1\,-\,\frac{\vert Q_c\vert}{\vert Q_h\vert}.
\end{equation}
The heat exchanged during the hot isochore $Q_h$ is the same for both the coherent and incoherent cycles: 
\begin{equation}
    Q_h\,=\,\frac{\omega_h}{2}\,\left(\tanh\left(\frac{\beta_c\,\omega_c}{2}\right)\,-\,\tanh\left(\frac{\beta_h\,\omega_h}{2}\right)\right).
    \label{eq:Qh}
\end{equation}

On the other hand, the heat exchange along the cold isochore $Q_c$ depends on the type of cycle and therefore is the one that makes the efficiency different for coherent and incoherent cycles. In order to calculate $Q_c$ we split the cold isochore in two steps. First, we calculate the heat exchange during the time that the hamiltonian is proportional to $\sigma_x$, and then we add the heat contribution when the hamiltonian is proportional to $\sigma_z$.
The heat rate, defined as $\dot{Q}_c\,=\,-\langle H_\mathcal{R}\rangle$ is given by:
\begin{equation}
    \dot{Q}\,=\,\omega_c\left(\gamma^+\,\langle\sigma^+\sigma^-\rangle\,-\,\gamma^-\,\langle\sigma^-\sigma^+\rangle\right),
\end{equation}
where $\gamma^\pm = \epsilon^2\frac{e^{\pm\beta_c \omega_c/2}}{Z_c}$ are the dissipation rates of the master equation that governs the dynamics of the system, and the average is taken with respect to the system's state. Thus,
\begin{equation}
    \dot{Q}_c\,=\,\omega_c\left(\gamma^+\rho_{00}(t)\,-\,\gamma^-\rho_{11}(t)\right),
    \nonumber
\end{equation}
and using that $\rho_{00}+\rho_{11}=1$ we have
\begin{equation}
    \dot{Q}_c\,=\,\omega_c\left(\gamma^+\,+\,\gamma^-\right)\rho_{00}(t)\,-\,\omega_c\gamma^-.
    \nonumber
\end{equation}
Finally, the total amount of heat exchanged is:
\begin{eqnarray}
   Q_c &=& \omega_c(\gamma^+\,+\,\gamma^-)\,\int_0^{t_d}\rho_{00}^x(t)\,dt\,+\nonumber \\
   &+& \omega_c (\gamma^+ + \gamma^-) \int_{t_d}^\infty \rho_{00}^z\,dt
    -\int_0^\infty \omega_c \gamma^-\,dt
\label{Q_integral}
\end{eqnarray}
where $\rho_{00}^x(t)$ ($\rho_{00}^z(t)$) is the excited state population during the interval where the hamiltonian is proportional to $\sigma_x$ ($\sigma_z$). Recall that the hamiltonian is changed at time $t_d$. Now, we can obtain the explicit evolution of each term from the master equation:
\begin{equation}
    \rho_{00}^z(t)\,=\,\left(\rho_{00}^x(t_d)\,-\,\frac{e^{\beta_c\,\omega_c/2}}{Z_c}\right)\,e^{-\epsilon^2\,t}\,+\,\frac{\gamma^-}{\epsilon^2},
\end{equation}
where $\rho_{00}^x(t_d)$ is the population of the excited state after the first step finishes. Inserting this into Eq.\eqref{Q_integral} gives
\begin{multline}
    Q_c\,=\,\omega_c\,\epsilon^2\,\int_0^{t_d}\,\rho_{00}^x(t)\,dt\,+ \\ +\,\omega_c\,\left(\rho_{00}^x(t_d)\,-\,\frac{e^{\beta_c\,\omega_c/2}}{Z_c}\right)e^{-\epsilon^2\,t_d}\,
    -\,\omega_c\,\gamma^-\,t_d.
    \nonumber
\end{multline}
To perform the last integration we need to know $\rho_{00}^x(t)$, which is obtained by solving the master equation with the system's hamiltonian proportional to $\sigma_x$:
\begin{equation}
    \rho_{00}^x(t)\,=\,B\,e^{-\frac{3}{8}\epsilon^2\,t}\,\cos(\Omega\,t)\,+\,\frac{4\omega_c^2\,-\,\epsilon^4\,e^{\beta_c\,\omega_c/2}/Z_c}{8\,\omega_c^2\,+\,e^4}
    \label{eq:rho_x}, \nonumber
\end{equation}
where
\begin{equation}
    B\,=\,\frac{e^{-\beta_h\,\omega_h/2}}{Z_h}\,-\,\frac{4h^{2}\,-\,\epsilon^{4}\,e^{\beta_c\,\omega_c/2}/Z_c}{8h^{2}+\epsilon^{4}},
    \nonumber
\end{equation}
and
\begin{equation}
    \Omega\,=\,\sqrt{\omega_c^{2}\,-\,\frac{\epsilon^{4}}{64}}.
    \label{eq:omega}
\end{equation}
Using this expression and performing the integration the heat exchanged $Q_c$ is equal to:
\begin{eqnarray}
    Q_c &=& 8\,\omega_c\,\epsilon^{2}Be^{-3/8\,\epsilon^{2}t_{d}}\left(\frac{8\Omega\,\sin\text{\ensuremath{\Omega t_{d}\,-\,3\epsilon^{2}\cos\Omega t_{d}}}}{9\epsilon^{4}+64\Omega^{2}}\right)+  \nonumber \\
    &+& \frac{24\epsilon^{4}\omega_c B}{9\epsilon^{4}+64\Omega^{2}} \nonumber \\
    &+&\left (\frac{4\omega_c^{2}\epsilon^{2}-\epsilon^{6}e^{\beta_c\omega_c/2}/Z_c}{8\omega_c^{2}+\epsilon^{4}}-\frac{e^{\beta_c\omega_c/2}}{Z_c}\right ) \omega_c t_{d} + \nonumber \\
    &-& h\left(\frac{e^{\beta_c\omega_c/2}}{Z_c}-Be^{-3/8\epsilon^{2}t_{d}}\cos\Omega t_{d}-\frac{4h^{2}-\epsilon^{4}e^{\beta_c\omega_c/2}/Z_c}{8h^{2}+\epsilon^{4}}\right) \nonumber \\ \nonumber 
\end{eqnarray}

Now we consider the weak coupling limit $\epsilon^2<<\omega_c$, that is the regime in which the coherent process shows the better performance. So we expand the last equation to first order in $\epsilon^2/\omega_c$. Under this approximation we have $\Omega\approx \omega_c$, $t_d\approx \pi/\omega_c$ and $B\approx \frac{e^{-\beta_h \omega_h/2}}{Z_h}-\frac{1}{2}$, and we get the following expression for the heat:
\begin{eqnarray}
Q_{c}&=&\omega_{c}(\tanh(\beta_{h}\omega_{h}/2)\,-\,\tanh(\beta_{c}\omega_{c}/2)) + \nonumber \\
    &+&\omega_{c}\left (\frac{e^{-\beta_{h}\omega_{h}/2}}{Z_{h}}\,-\,\frac{e^{-\beta_{c}\omega_{c}/2}}{Z_{c}}\right)+\frac{5}{8}\epsilon^{2}\pi\left(\frac{1}{2}-\frac{e^{\beta_{c}\omega_{c}/2}}{Z_{c}}\right). \nonumber
\end{eqnarray}

Therefore, recalling that $Q_h$ is given by Eq.~\eqref{eq:Qh}, we can express the efficiency of the coherent cycle in the weak coupling limit as:
\begin{equation}
    \eta\,=\,\eta_{Otto}-\omega_{c}\frac{(\frac{e^{-\beta_{h}\omega_{h}/2}}{Z_{h}}\,-\,\frac{e^{-\beta_{c}\omega_{c}/2}}{Z_{c}})\,-\,\frac{5}{8}\frac{\epsilon^{2}}{\omega_c}\pi(\frac{e^{\beta_{c}\omega_{c}/2}}{Z_{c}}\,-\,\frac{1}{2})}{\vert Q_{h}\vert}.\nonumber
\end{equation}
Since in this limit $\frac{\epsilon^2}{\omega_c}<<1$ the efficiency of the coherent cycle is upper bounded by the Otto efficiency.


%

\end{document}